\documentstyle[12pt,a4,cite,graphicx]{article}

\newcommand{\ba}{\begin{eqnarray}}
\newcommand{\ea}{\end{eqnarray}}
\newcommand{\be}{\begin{equation}}
\newcommand{\ee}{\end{equation}}

\begin{document}

\begin{titlepage}
\begin{flushright}
CERN-PH-TH/2007-108\\
\end{flushright}
\vspace{2cm}
\begin{center}

{\large\bf ChPT Progress on Non-Leptonic and Radiative Kaon Decays
\footnote{Invited talk at ``KAON '07 
International Conference'', May 21-25  2007, Frascati, Italy.}}\\
\vfill
{\bf  Joaquim Prades}\\[0.5cm]

Theory Unit, Physics Department, CERN  
CH-1211  Gen\`eve  23, Switzerland \\
and CAFPE and Departamento de
 F\'{\i}sica Te\'orica y del Cosmos, Universidad de Granada, 
Campus de Fuente Nueva, E-18002 Granada, Spain.\\[0.5cm]

\end{center}
\vfill

\begin{abstract}
\noindent
I discuss recent developments
 on non-leptonic and radiative kaon decays
mainly related to direct CP-violation
within the combined ChPT and $1/N_c$ expansion approaches.
In particular, I review the status of $K \to \pi \pi$,
$\varepsilon_K'$, direct CP-violating $K^+ \to 3 \pi$ 
Dalitz plot slope $g$ and decay rate asymmetries, and the Standard Model
prediction for $Br(K_L \to \pi^0 e^+ e^-$).
\end{abstract}
\vfill
July 2007
\end{titlepage}
\setcounter{page}{1}
\setcounter{footnote}{0}

\section{Motivation}
Non-leptonic  and radiative kaon decays have attracted
a lot of attention in various respects.  
Testing the Standard Model (SM)
and unveiling flavour structure beyond it is one of them.
This can be done  very effectively using
precision tests of the scalar  sector where
{\em direct} CP-violating effects involving kaons provides with
some of the most promising opportunities.
Indeed, {\em direct} CP violation in kaon  decays
is experimentally very well known in $K\to \pi \pi$ \cite{epsilonp,SOZ04}
\be
{\rm Re} \left( \varepsilon_K' / \varepsilon_K \right)
= \left( 1.63 \pm 0.23 \right) \times 10^{-3}.
\label{epsp}
\ee
 I discuss the present theoretical status of the SM prediction
for this quantity in Section \ref{K2pi}  while
CP-violating $K^+ \to 3 \pi$ Dalitz plot slope $g$ and decay
rate asymmetries are in  Section \ref{K3pi} .

As a typical example of radiative kaon decays,
I discuss in Section \ref{rad} the theoretical advances
predicting the CP-violating decay
$K_L \to \pi^0 \gamma^* \to \pi^0 e^+ e^-$ within the SM.

A deeper understanding of the strong-weak dynamics interplay
at low energy is also a very interesting aspect of  the study of
non-leptonic and radiative kaon decays.
Finally, I also report on the  recent theoretical advances
based  in large $N_c$  approaches to low-energy QCD.

\section{Theoretical Framework}
\label{framework}

The SM effective action  at energies  around or below the charm
quark mass is well known. For the $\Delta S=1$
sector, this  has been done to next-to-leading order 
(NLO) in two renormalization
schemes (NDR and HV) by two groups, \cite{BBL96} and 
\cite{CIU95}. It contains ten four-quark operators, $Q_1$ to $Q_{10}$,  
and two  magnetic dipole operators, $Q_{11}$ and $Q_{12}$,
which are chirally suppressed, see e.g. \cite{BBL96} for definitions. 
  In the presence of electroweak interactions, 
there appear another two operators, $Q_{7V}$ and $Q_{7A}$, 
 which contribute to radiative kaon decays, see e.g. 
\cite{BBL96}. Short-distance information enters
via Wilson coefficients multiplying the operators of the effective
action. This short-distance information
is the one  we want to extract from measurements of 
non-leptonic and radiative kaon decays.

 For the explicit expression of the $\Delta S=1$ SM effective action
 and a very detailed discussion 
of low-energy SM effective action see \cite{BBL96}.
Here I use the same notation as there.

Chiral Perturbation Theory (ChPT) \cite{WEI79,GL84} is the 
effective field theory that describes the SM interactions 
among  the lowest-energy 
degrees of freedom: pions, kaons, photons, $\cdots$
 For reviews with emphasis on  kaon physics see \cite{chiReviews}.
   There have been  recent advances and a lot of
work  in understanding the long-distance--short-distance
matching between the effective SM action and ChPT, 
both  using analytical large $N_c$ methods and 
lattice QCD -- for lattice, 
see Chris Sachrajda and Bob Mawhinney's talks.  As yet, 
there remains a lot of work to be done,
mainly for rare kaon decays. 

Within ChPT, one constructs the most general Lagrangian compatible
with all SM  symmetries and  in particular, with the structure generated
 by the QCD chiral symmetry breaking SU(3)$_L$ $\times$ SU(3)$_R$ $\to$
SU(3)$_V$.  ChPT provides  then with  a low-energy
Taylor expansion of amplitudes in external momenta and meson masses
which  in general depends on unknown couplings.
This is still very predictive because,
 at lower orders, there appear only few of them, e.g.  
just the pion $\pi^+ \to \mu^+ \nu$
decay constant in the chiral limit, $F_0$, 
and  the lowest pseudo-Goldstone boson octet  masses in the 
strong sector at leading-order (LO).
This fact  allows to relate different decays with the same unknowns. 
 Also in SU(3) but at next-to-leading order (NLO) and 
without electromagnetism (EM), ten additional 
physical couplings, $L_1$ to $L_{10}$,
 \cite{GL84} are needed in the $\Delta S=0$ sector. 
One more coupling appears when including EM at LO.

In other cases, it can be shown that LO
 chiral loops are finite and no unknown counterterm 
at that order appears --these are parameter free
predictions at that order. To this class  belong the   
radiative decays $K_S \to \gamma \gamma$ \cite{AE86}   and
 $K_L \to \pi^0   \gamma \gamma$ \cite{EPR87a}.  For both decays there
 have been reported new measurements at this Conference.
In the case of $K_S \to \gamma \gamma$, the KLOE result  \cite{MAR07}
nicely confirms the LO ChPT prediction 
 while the KTeV preliminary result
\cite{CHE07} agrees with a previous NA48 measurement
pointing to the  need of large NLO  ChPT corrections.
For a complete  discussion of these two decays 
and for a comprehensive list of works  applying ChPT to non-leptonic
and rare kaon decays see \cite{AEI94}.

At LO in the  $|\Delta S|=1$ SM sector  and within SU(3), 
there appear three  couplings \footnote{There appears one more
octet singlet coupling within U(3), see \cite{GTS05}.} 
 of order $p^2$  plus one of order $e^2 p^0$, namely,
$G_8$, $G_8'$, $G_{27}$ and $G_E$, respectively.
The corresponding Lagrangian reads
\ba
{\cal L}^{(2)}_{|\Delta S|=1} &=&
C F_0^6 \, e^2 G_E \, {\rm tr} \left( \Delta_{32} u^\dagger Q u \right) 
+ C F_0^4 \, \left[ G_8 {\rm tr} \left( \Delta_{32} u_\mu u^\mu \right) 
\right. 
\nonumber \\ &+& \left. G_8' \, {\rm tr} \left( \Delta_{32} \chi_+ \right)
+ G_{27} \, t^{ij,kl} {\rm tr} \left( \Delta_{ij} u_\mu \right) 
{\rm tr} \left( \Delta_{kl} u^\mu \right) \right]
\ea
with $C=-(3/5) G_F V_{ud} V_{us}^* / \sqrt 2 \simeq -1.08 \times 10^{-6}
\, {\rm GeV}^{-2}$, 
$u_\mu \equiv i u^\dagger (D_\mu U) u^\dagger$, $U\equiv u^2 =
\exp (i \sqrt 2 \Phi /F_0)$, $\Delta_{ij}= u \lambda_{ij} u^\dagger$,
$(\lambda_{ij})_{ab} = \delta_{ia} \delta_{jb}$, $\chi_+=
u^\dagger \chi u^\dagger + u \chi^\dagger u$, $\chi={\rm diag}
(m_u, m_d, m_s)$ and the $t^{ij,kl}$ tensor can be found 
in \cite{BPP98}.
The SU(3) $\times$ SU(3) matrix $\Phi$ collects pion, kaon and eta
pseudo-Goldstone boson fields.
In this normalization, $G_8=G_{27}=1$ at large $N_c$.
At NLO in ChPT,  the $|\Delta S|=1$ SM sector was constructed 
within SU(3) in \cite{EPR87a,couplings,EPR87b}.

\section{Non-Leptonic Kaon Decays}

\subsection{$K\to \pi \pi$ and $\varepsilon_K'$: Status}
\label{K2pi}

  The decays $K \to \pi \pi$ are fully known to NLO in ChPT,
i.e. including isospin breaking effects from quark masses
and EM  \cite{BPP98,KMW90,PPS00,CG00,CEI04}. The r\^ole of final
state interactions (FSI) in those decays is also clarified.
For a recent summary of the  theory status of both
the $\Delta I=1/2$ rule in kaons and  $\varepsilon_K'$ see 
\cite{BGP04}.

In \cite{BB03}, the authors performed a combined fit  to 
 both  data on $K \to \pi \pi$ and $K \to 3 \pi$ which is also
known fully at NLO in ChPT including isospin breaking 
\cite{BB03,GPS03} and obtained
\be
{\rm Re} \, G_8 = (7.0 \pm 0.6) \left( 87 \, {\rm MeV} / F_0 \right)^4
\, ;  \, \, 
G_{27} = (0.50 \pm 0.06) \left( 87 \, {\rm MeV} / F_0 \right)^4
\ee
 which represents  the $\Delta I= 1/2$ rule for kaons.
Recent analytical advances on the quantitative understanding of this
rule can be found in \cite{BP99,HPR03} using $1/N_c$ approaches. 
In particular, the
$\Delta I=1/2$ rule is reproduced within 40 \% in \cite{BGP04,BP99}
at NLO in $1/N_c$ using the ENJL model \cite{ENJL} at low energies
  and with analytical short-distance independence.

Using the calculations quoted above, one can get the prediction for 
$\varepsilon'_K$  fully at NLO in ChPT
\be
{\rm Re} \left( \varepsilon_K'/\varepsilon_K\right)
\simeq - \left[\left(1.9\pm0.5 \right) \, {\rm Im}
G_8 + \left(0.34\pm0.15 \right) {\rm Im}
(e^2 G_E) \right] \, 
\label{CHepsp}
\ee
where ${\rm Im} \, G_8$ and ${\rm Im}\, (e^2 G_E)$ are proportional to the
CP-violating phase ${\rm Im} \, \tau \equiv - {\rm Im} \, 
(\lambda_t/\lambda_u)$ with $\lambda_i \equiv  V_{id} V_{is}^*$
and $V_{ij}$ are Cabibbo--Kobayashi--Maskawa matrix elements.
Notice that  it does not appear
any $p^4$ counterterm ${\rm Im} \, \widetilde K_i$ --see \cite{BB03} 
for their definition-- in the previous NLO in ChPT expression 
for ${\rm Re} \, (\varepsilon_K'/\varepsilon_K)$ because  they 
have been estimated to be negligible within large $N_c$ \cite{PPS00}.

\begin{figure}
\center{\begin{minipage}[t]{0.8cm}\vskip2.cm 
\large{$\frac{{\rm Im} G_8}{{\rm Im} \tau}$}\end{minipage}
\rotatebox{-90}
{\includegraphics[width=5.cm]{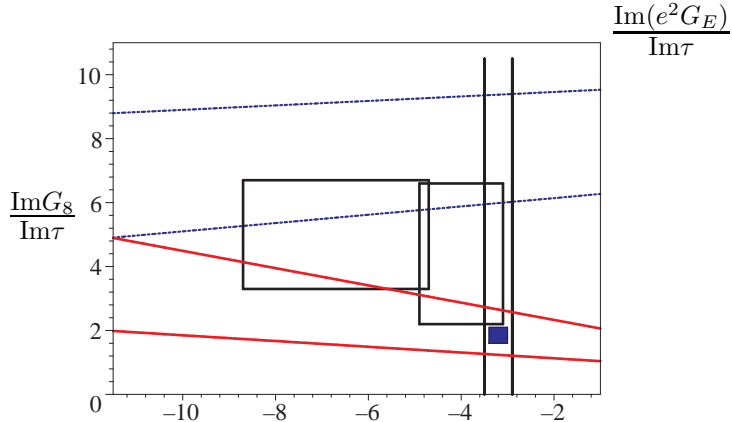}}
\begin{minipage}[b]{1cm}\hspace*{-0.5cm}\vspace{-5.5cm}
\large{$\frac{{\rm Im} (e^2 G_E)}{{\rm Im} \tau}$}\end{minipage}}
\vspace*{0.2cm}
\caption{$\varepsilon_K'$: Theory vs Experiment. See text for explanation.}
\label{fig:status}
\end{figure}

Putting together the experimental result in (\ref{epsp}) and the
NLO ChPT formula in (\ref{CHepsp}), one obtains that 
the pair (${\rm Im} \, (e^2 G_E)$, ${\rm Im} \, G_8$)
has to lie between  the two lower horizontal (red) lines in Fig. 
\ref{fig:status}. An immediate consequence of
(\ref{CHepsp}) is  that for typical values
of ${\rm Im} \, (e^2 G_E)$ and ${\rm Im} \, G_8$ 
--say large $N_c$ values--   though there is some cancellation
between the two terms there, it is however not as large
as it was  previously thought and still sometimes argued.

 Recently, several  analytical works have been devoted to calculating
${\rm Im} \, (e^2 G_E)$ \cite{CDH01,BGP01,FGR04,NAR01}  --see \cite{FGR04} 
for a comparison--  and
${\rm Im} \, G_8$ \cite{HPR03,BP00}, both at NLO in the $1/N_c$.
The nice feature of ${\rm Im} \, (e^2 G_E)$  is that it can be related
via dispersion relations to  $VV-AA$ spectral two-point  
function in the chiral limit \cite{CDH01,BGP01,NAR01}. 
The results found for the pair  
(${\rm Im} \, (e^2 G_E)$, ${\rm Im} \, G_8$) in 
\cite{BGP01,BP00} are represented in Fig. \ref{fig:status} 
by the rectangle on the right
while the results  in \cite{HPR03,FGR04} are represented by the
rectangle on the left.  In these two calculations, part of the large 
uncertainties  come from  two input parameters, namely, 
the quark condensate in the chiral limit which present
 uncertainty is around 20 \% and  enters squared 
 and $L_5$ which uncertainty is around 45 \%. 
The large $N_c$ result is the  (blue) filled square 
to which I have not assigned any uncertainty since it does not include
the NLO in $1/N_c$ different planar topology.  
 The lattice result for ${\rm Im} \, (e^2 G_E)$ \cite{lattice} 
is also shown in Fig. \ref{fig:status}, 
it lies between the two vertical lines. Unfortunately,
we still don't have a reliable value for ${\rm Im} \, G_8$ from lattice
QCD but one can assess from Fig. \ref{fig:status} what that value
has to be if compatible with the  measurement of  $\varepsilon_K'$.

\subsection{$K \to 3 \pi$ and Direct CP-Violating Dalitz-Plot Slope
$g$ Asymmetries}
\label{K3pi}

Non-leptonic $K\to 3 \pi$ decays have also attracted a lot of work 
recently. Theses decays were calculated at NLO in ChPT in \cite{KMW90}
but unfortunately the complete expressions were not available
 and as mentioned above, recently they were redone in \cite{BB03,GPS03}. 
Using those calculations, one can  predict the Dalitz plot slopes
--see e.g. \cite{GPS03} for their definition-- at NLO in ChPT
for $K^+ \to \pi^+ \pi^+ \pi^-$ and $K^+ \to \pi^0 \pi^0 \pi^+$ 
which are in very good agreement with  recent measurements
 \cite{slopes}.

It is possible to define CP-violating asymmetries 
using the Dalitz plot slope $g$ \cite{GPS03,previous}. Previous
predictions within the SM were done using LO ChPT
plus various NLO estimates \cite{previous}. There  is
work  looking for large SUSY effects in  this asymmetries
as well \cite{AIM00}.  The first full NLO in ChPT results
were presented in \cite{GPS03} where one can also find 
 the $K^+ \to 3 \pi$ decay rate CP-violating asymmetries.
At NLO in ChPT in the isospin limit, one gets
for the $K^+ \to \pi^+ \pi^+ \pi^-$ slope $g_C$
 \ba
10^2 \times \Delta g_C &\simeq&
(0.7\pm 0.1) \, {\rm Im} \, G_8 - (0.07\pm 0.02) \, {\rm Im}
\, (e^2 G_E) \nonumber \\ &+& (4.3\pm1.6) \, {\rm Im} \, \widetilde K_2 -
(18.1\pm2.2)\, {\rm Im} \, \widetilde K_3 \, 
\ea
where $\widetilde K_i$ are order $p^4$ counterterms
\cite{BB03,GPS03}. 
More details and similar expressions for $\Delta g_N$ and the decay 
rate asymmetries can be found in \cite{GPS03}. It turns out that 
$\Delta g_C$ is quite stable against
unknown  NLO ChPT counterterms while $\Delta g_N$ 
is  somewhat less stable \cite{GPS03}. 
The results obtained are \cite{GPS03}
\be
\Delta g_C=-(2.4\pm1.2) \times 10^{-5}; \, \, \, 
\Delta g_N=(1.1\pm0.7) \times 10^{-5} \,.
\ee
 Variation of input values and other uncertainties
are within the quoted  error.
Experimentally, the final results of the NA48/2 experiment
were presented at this Conference \cite{NA48asym},
\be
\Delta g_C=-(1.5\pm 2.1) \times 10^{-4}; \, \, \, 
\Delta g_N=(1.8\pm1.8) \times 10^{-4} \, .
\ee
which are compatible with previous measurements \cite{ISTRAasym} but with
significantly smaller uncertainty.

In Fig. \ref{fig:status}, the region between 
the two upper horizontal (blue) 
lines is where the pair (${\rm Im} \, (e^2 G_E)$, ${\rm Im} \, G_8$)
 would have to lie if $\Delta g_C = -(4.0\pm0.5) \times 10^{-5}$  
was measured. Any measurement of  $\Delta g_C$  between this value
and the present experimental limits would lead to an allowed region
for the pair  (${\rm Im} \, (e^2 G_E)$, ${\rm Im} \, G_8$)  which moves
toward the upper side of that figure when the modulus of $\Delta g_C$
increases. Therefore, if we require this region
to cross with the  allowed region for $\varepsilon_K'$ then we  
would need a  negative very large value in modulus 
for  ${\rm Im} \, (e^2 G_E)/ {\rm Im} \tau$. 
Using the results from  calculations
of this coupling both using  analytic techniques  
\cite{CDH01,BGP01,FGR04,NAR01}
and lattice QCD --see Chris Sachrajda and Bob Mawhinney's talks 
at this Conference,  this  would clearly call for 
the presence of new physics independently of the hadronic uncertainties
in  ${\rm Im} \, G_8$. 
This plot also points to an experimental accuracy in $\Delta g_C$
of around $0.2 \times 10^{-4}$  as the goal to be reached.

\section{Radiative Kaon Decays}
\label{rad}

As a typical example of radiative kaon decay, I discuss here
the status  and make some comments on the $K\to \pi \gamma^* \to
\pi \ell^+ \ell^-$ decays. ChPT at LO plus  NLO dominant effects  analysis
have been done and unknown couplings appear \cite{EPR87b,EDI98}.
The short-distance contribution to  the SM effective action 
description is  also known  at  NLO order in two 
schemes (HV and NDR) \cite{BBL96,GW80,BLM94}.
On the experimental side,  the CP-conserving $K^+ \to \pi^+ \ell^+ \ell^-$
and $K_S \to \pi^0 \ell^+ \ell^-$, which are dominated by the
long distance process $K \to \pi \gamma^* \to \pi \ell^+ \ell^-$,
 have been measured.

At LO in ChPT a single coupling governs the 
$K \to \pi \gamma^*$ form factor \cite{EPR87b}.  In the case of
$K^+ \to \pi^+ \ell^+ \ell^-$ this coupling is of order $N_c$
and was called $\omega_+$.
 The authors of \cite{EDI98} pointed out that adding 
 a NLO momenta dependent term to the form factor
 improves considerably the fit. Including  this NLO term and 
using the measurement at  BNL \cite{BNL99},
one gets ${\rm Re} \, \omega_{+,e}=1.49\pm 0.02$ or equivalently, 
${\rm sign}(G_8) \, a_{+,e}=-(0.59\pm0.01)$. Subscript 
$e$ refers to the electron mode, see \cite{EDI98}  for the definition
of $a_+$.
 The corresponding decay into muons has also been measured
giving compatible results \cite{HYP02}.
Notice that both $\omega_{+}$ and ${\rm sign}(G_8) \, a_{+}$
are global sign convention independent.

Analogously,  one can obtain the  coupling
that governs the $K^0 \to \pi^0 \gamma^*$ form factor 
at LO from the measurement of $K_S \to \pi^0 e^+ e^-$, 
in this case this  coupling
is of order one in $1/N_c$ and was called $\omega_S$.
The different  $N_c$ counting of $\omega_+$ and $\omega_S$
already tells us that they are unrelated 
as noticed in \cite{BP93,DG95}.  In this case, 
NLO momenta dependent  terms in the form factor cannot be
determined from a fit to data due to the smallness of the
non-analytic contributions \cite{DBI03}. The result one gets 
using the NA48/1 results \cite{BAT03} has a twofold ambiguity
 ${\rm Re} \, 
\omega_{S,e}=[2.53^{+0.56}_{-0.45}, \, -(1.87^{+0.56}_{-0.45})]$ 
which does not fix the sign of the coupling.
Equivalently, using the notation of \cite{EDI98} one gets
$|a_{S,e}|=1.12^{+0.29}_{-0.23}$. The corresponding decay into 
muons has also been measured giving compatible results \cite{BAT04}.

The closely related CP-violating $K_L \to \pi^0 \ell^+ \ell^-$ decay
  has received a great deal of attention both within the SM 
\cite{EPR87b,EDI98,GW80,BLM94,DG95,DBI03,ISU04,FGR05,MTS06,MS07}
and as tool of unveiling  beyond the SM flavour structure \cite{BLA07}.
A pretty precise prediction for this decay within the SM can be made.
In particular, it was shown in \cite{DBI03,ISU04} that the 
CP-conserving $K_L \to \pi^0 \gamma^* \gamma^* \to \pi^0 e^+ e^-$
decay contribution is negligible. 
Updating \cite{DBI03,MTS06} and using \cite{MS07}, one gets
\ba
Br(K_L \to \pi^0 e^+ e^-) &=&
\left[ (3.41 \pm 0.03) \, W_{S,e}^2 +
(3.91\pm 0.05) \, W_{S,e} \,  (\widehat y_{7V}+ M_{6V}) \, \left(
\frac{{\rm Im} \lambda_t}{10^{-4}} \right) 
\nonumber \right. \\ 
&+&  \left. (2.36\pm0.06) \, \left[ \widehat y_{7A}^2 + (\widehat y_{7V}+
M_{6V})^2 \right] \, \left(\frac{{\rm Im} \lambda_t}{10^{-4}} \right)^2 
\right] \times 10^{-12}
\label{KLCP}
\ea
with $W_{S,e}^2 \equiv 10^9 \times Br(K_S \to \pi^0 e^+ e^-)/1.20$
and,  to a very good approximation,  $W_{S,e}= {\rm Re} \, 
\omega_{S,e}- 1/3$, and $\widehat y_{7V(A)} \equiv  y_{7V(A)}/\alpha$ 
\cite{BBL96}. The term $M_{6V}$ is the hadronic
penguin operator $Q_6$ contribution to the direct CP-violating term.
The $Q_{7V}$ relevant matrix element is $3/4\pi\alpha$  and
the $Q_6$ one  is, at large $N_c$,
\be
\langle Q_6 \rangle {\big|}_{N_c} (\nu)
= 32 \frac{\langle \overline q q \rangle^2(\nu)}{F_0^6} \, 
\left[ 2 C_{63}^r - C_{65}^r\right](M_\rho) \, .
\label{Q6}
\ee
 where $C_{63}^r$ and $C_{65}^r$ are two $\Delta S=0$ 
${\cal O}(p^6)$ couplings \cite{BCE99}.  
This same  combination of counterterms appears in  
the EM $K^0$ charge radius NLO ChPT calculation \cite{BT02}. Using  
the PDG \cite{PDG06,CHARGE}  experimental value, one gets 
\be
\left[2 C_{63}^r - C_{65}^r\right][M_\rho]=
(1.8\pm 0.7)\,(F_0 / 87 \, {\rm MeV})^2 \,  \times 10^{-5}  \, 
\label{exp}
\ee
which together with (\ref{Q6}) yields
\be
\frac{M_{6V}}{\widehat y_{7V}}\equiv
\frac{y_6(\nu) \, \langle Q_6 \rangle(\nu)}{y_{7V}(\nu) 
\, \langle Q_{7V} \rangle }
= -(0.2\pm 0.1) \, B_{6V}
\ee
where $B_{6V}$  parameterizes non-factorisable corrections.
This contribution, which has been argued before  to be negligible 
\cite{BBL96,BLM94,DBI03}, adds to the direct CP-violating 
vector part  and could  be as large as $-(30\sim 50)$ \% of the $Q_{7V}$
contribution depending of the unknown  $B_{6V}$ factor.

The interference  term  in (\ref{KLCP})
is constructive (destructive) if  ${\rm Re} \, \omega_{S,e}$ 
is larger (smaller) than $1/3$. 
Or equivalently, if ${\rm sign} (G_8) \, a_{S,e}$ is
positive (negative).
The $Q_{7V}$ contribution   is model independent and gives
${\rm sign} (G_8) \, a_S^{Q_{7V}} > 0$,
 i.e. constructive interference  \cite{GW80}.  Assuming  VMD  for the 
$K_S\to \pi^0 \gamma^*$ form factor and a large non-VMD contribution
for the $K^+ \to \pi^+ \gamma^*$ plus the $Q_{7V}$ relation
$ a_S^{Q_{7V}, VMD} = - a_+^{Q_{7V}, VMD}$  produces
${\rm sign} (G_8) \, a_S >0$ and therefore constructive
interference  if one  furthermore identifies $a_S^{Q_{7V}, VMD}$  and
$a_+^{Q_{7V}, VMD}$ with the experimental values 
for $a_S$ and $a_+$, respectively \cite{DBI03}. This
identification is not trivial  as these couplings receive
 sizable contributions from the hadronic operators
$Q_2$ and $Q_6$ which do not fulfill the above $Q_{7V}$ relation between
$a_S$ and $a_+$ \cite{BP93,PRA07}.

In \cite{FGR05}, the authors saturated  $K \to \pi \gamma^*$
form factor by $K^*$ and $\rho$  meson single poles
 within a large $N_c$ inspired minimal hadronic
approximation which used to make 
a fit to  data. They got ${\rm Re} \, 
\omega_+=1.4\pm 0.6 > 1/3$  [i.e. ${\rm sign} (G_8) \, a_+ =
-(0.5\pm0.3) <0$] and  ${\rm Re} \, \omega_S=-(2.1\pm 0.2) < 1/3$  
[i.e. ${\rm sign} (G_8) \, a_S =-(1.2\pm0.1) <0$] which implies 
  destructive  interference \footnote{Notice that the sign
of the interference term in (\ref{KLCP}) agrees  with 
\cite{EDI98,GW80,DBI03} but is opposite to that used in \cite{FGR05}.}.

In  \cite{BP93}, in addition to the contribution of $Q_{7V}$,
a four-quark effective action model was used to calculate
the contributions from  $Q_{i=1, \cdots, 6}$.
These authors \underline{predicted}
 ${\rm Re} \, \omega_+=1.5^{+1.2}_{-0.6} > 1/3$  
[i.e. ${\rm sign} (G_8) \, a_+ = -(0.6^{+0.6}_{-0.3}) <0$] 
and  ${\rm Re} \, \omega_S=1.6^{+1.4}_{-0.6} > 1/3$  
[i.e. ${\rm sign} (G_8) \, a_S = 0.6^{+0.7}_{-0.3} > 0$]
which implies  constructive interference.
In particular,  the  large $N_c$ result for
$\langle Q_6 \rangle$  in \cite{BP93} is equivalent to
\be
\left[2 C_{63}^r - C_{65}^r\right][1 {\rm GeV}]
=(2.2\pm 1.1)\,(F_0 / 87 \, {\rm MeV})^2 \, 
 \times 10^{-5}  \, 
\ee
which compares well with (\ref{exp}).  
In fact, it is easy to see that $Q_6$ with the large $N_c$ 
result  in (\ref{Q6}) together with (\ref{exp})
contributes to ${\rm Re} \, \omega_S$ with 
the  same sign  as $Q_{7V}$ and comparable magnitude.
An analysis of the rest of contributions to $\omega_S$ from
four-quark operators can be performed at NLO in $1/N_c$  \cite{PRA07}
using the approaches developed in \cite{BGP01,BP00,KPR00}.

As pointed out in \cite{MTS06}, one can determine experimentally
the sign of the interference term in (\ref{KLCP})
 using the $K_L \to \pi^0 \mu^+ \mu^-$ forward-backward
asymmetry.  Actually, this study can also serve to fix
the long-distance contribution to the direct CP-violating
term $M_{6V}$ which has to be treated as a further unknown at present.
  Both, $K_L \to \pi^0 e^+ e^-$ and
$K_L \to \pi^0 \mu^+ \mu^-$ modes become then necessary to disentangle 
new physics \cite{MTS06}  and long-distance 
from short-distance  direct CP-violating terms.
Using  ${\rm Im} \, \lambda_t=(1.4 \pm 0.2)
\, \times \,  10^{-4}$ \cite{PDG06}, $\widehat y_{7A}=-(0.68\pm 0.03)$,
$\widehat y_{7V}=0.73\pm 0.04$ \cite{BBL96,BLM94} in (\ref{KLCP})
with  constructive [destructive] interference, one gets 
predictions for $Br(K_L\to \pi^0 e^+ e^-)$
between $(2.7^{+0.9}_{-0.7}) \, \times 10^{-11}$ and 
$(2.5^{+0.9}_{-0.7}) \, \times  10^{-11}$ 
[$(1.3^{+0.9}_{-0.7}) \, \times 10^{-11}$ and  
$(1.4^{+0.9}_{-0.7})\, \times 10^{-11}$],
 if one varies $B_{6V}$ between one and two.
 The  present experimental limit is $Br(K_L \to \pi^0 e^+ e^-) 
<2.8 \times 10^{-10}$ \cite{KTeV04}.

\subsection{Some Selected Topics}

 Here, I would like to comment very briefly on two selected topics.
First, the NA48/2 very recent first measurement of 
a \emph{destructive} direct electric emission interference in $K^+ \to
\pi^+ \pi^0 \gamma$ \cite{RAG07}.
This interference depends on the sign  of one unknown  ChPT coupling
\cite{Kpigamma} and naive theoretical predictions 
tend to tell that it is \emph{constructive} \cite{BP93,Kpigamma}. 
 Clearly, more theory work is needed here.

Secondly, interesting recent work  on 
U$_A(1)$ anomaly effects in radiative kaon decays
using U(3) ChPT was done in \cite{GTS05} reaching a better
understanding of $K_L \to \gamma \gamma$ and $K_L \to \gamma
\ell^+ \ell^-$. One of the conclusions
reached there is that  U$_A(1)$ anomaly effects  could be 
sizable in $K_S \to \pi^0 \gamma \gamma$ and 
$K^+ \to \pi^+ \gamma \gamma$ and more experimental input on these
modes is very welcome.

\section{Conclusions}

To reach the goals of non-leptonic and radiative kaon decays  
studies, i.e. to obtain new flavour structure  (CP-violating phases) 
information  and/or  understand  the strong-weak 
dynamics  interplay, one needs in general
to combine different modes  to disentangle SM from new physics
 and/or long-distance from short-distance effects. This strategy
is  both complementary and necessary, see for instance
\cite{GPS03,MTS06,BLA07} where the cases
$\varepsilon_K'$ vs  $K^+ \to 3\pi$ CP-violating
Dalitz plot slopes asymmetries, $K_L \to \pi^0 e^+ e^-$ vs
$K_L \to \pi^0 \mu^+ \mu^-$ and  $\varepsilon_K'$ vs 
$K_L \to \pi^0 e^+ e^-$  have been studied, respectively.

At the same time, it is obvious the need of theoretical effort 
predicting unknown ChPT couplings in order to take profit of  
high precision measurements such as $\varepsilon_K'$,
and eventual measurements of   the CP-violating Dalitz plot 
slopes asymmetries, $K_L  \to \pi^0 \ell^+ \ell^-$,
$\cdots$ as unique probes unveiling physics beyond the SM.
 For recent efforts in that direction using 
large $N_c$ hadronic approaches see \cite{KPR00,BGL03,CIR06}
and references therein.
Lattice proposals to study radiative kaon decays
also appeared \cite{IMT06} while I refer
to Chris Sachrajda and Bob Mawhinney's talks at this Conference
for non-leptonic  kaon decays lattice efforts.

As a final remark,  I believe that
with the expected theory and experimental efforts, non-leptonic and 
radiative kaon decays will continue   provide with very nice and 
interesting physics.


\begin{thebibliography}{99}
\bibitem{epsilonp}
 A. Alavi-Harati {\it et al.}  [KTeV Collaboration],
  \emph{Phys. Rev.  D} {\bf 67} (2003) 012005 
  [Erratum-ibid.\  {\bf 70} (2004) 079904]
{\tt [hep-ex/0208007]};
J.R. Batley {\it et al.}  [NA48 Collaboration],
  \emph{Phys. Lett.  B} {\bf 544} (2002) 97
{\tt [hep-ex/0208009]}
 G.D. Barr {\it et al.}  [NA31 Collaboration],
  \emph{Phys. Lett.  B} {\bf 317} (1993) 233;
L.K. Gibbons {\it et al.},
  \emph{Phys. Rev. Lett.}  {\bf 70} (1993) 1203.

\bibitem{SOZ04}
 M.S. Sozzi,  \emph{Eur. Phys. J.  C} {\bf 36} (2004) 37
 {\tt [hep-ph/0401176]}.

\bibitem{BBL96}
G. Buchalla, A.J. Buras and M.E. Lautenbacher,
  \emph{Rev.\ Mod.\ Phys.}  {\bf 68} (1996) 1125
{\tt [hep-ph/9512380]}
and references therein.

\bibitem{CIU95}
M. Ciuchini, {\it et al.}, \emph{Z. Phys. C} {\bf 68} (1995) 239
{\tt [hep-ph/9501265]}.

\bibitem{WEI79}
S. Weinberg,  \emph{Physica A} {\bf 96} (1979) 327.

\bibitem{GL84}
J. Gasser and H. Leutwyler,  
\emph{Annals Phys. (NY)}  {\bf 158} (1984) 142;
\emph{Nucl.\ Phys.\  B} {\bf 250} (1985) 465.

\bibitem{chiReviews}
G. Ecker,   \emph{Prog.\ Part.\ Nucl.\ Phys.}  {\bf 35} (1995) 1
{\tt [hep-ph/9511412]};
E. de Rafael,  {\tt hep-ph/9502254};
A. Pich, \emph{Rept.\ Prog.\ Phys.}  {\bf 58} (1995) 563
{\tt  [hep-ph/9502366]}.

\bibitem{AE86}
G. D'Ambrosio and D. Espriu, \emph{Phys. Lett B} {\bf 175} (1986) 237;
J.L. Goity, \emph{Z. Phys. C} {\bf 34} 341 (1987).

\bibitem{EPR87a}
G. Ecker, A. Pich and E. de Rafael, 
\emph{Phys. Lett. B} {\bf 189} (1987) 363.

\bibitem{MAR07}
M. Martini, [KLOE Collaboration], these proceedings.

\bibitem{CHE07}
E. Cheu, [KTeV Collaboration], these proceedings.

\bibitem{AEI94}
 G. D'Ambrosio, G. Ecker, G. Isidori, and H. Neufeld,
  {\tt hep-ph/9411439};
G. D'Ambrosio and G. Isidori,
  \emph{Int.\ J.\ Mod.\ Phys.\  A} {\bf 13} (1998) 1
{\tt [hep-ph/9611284]};
G. D'Ambrosio,
\emph{Nucl.\ Phys.\ B (Proc.\ Suppl.)}  {\bf 66} (1998) 482
{\tt [hep-ph/9709314]};
{\tt hep-ph/0110354};
  \emph{Mod.\ Phys.\ Lett.\  A} {\bf 18} (2003) 1273
   {\tt [hep-ph/0305249]}.


\bibitem{GTS05}
J.-M. G\'erard, S. Trine and C. Smith,
\emph{Nucl.\ Phys.\  B} {\bf 730} (2005) 1
 {\tt [hep-ph/0508189]}.

\bibitem{BPP98}
J. Bijnens, E. Pallante and J. Prades,
 \emph{Nucl.\ Phys.\  B} {\bf 521} (1998) 305
{\tt [hep-ph/9801326]}.

\bibitem{couplings}
G. Ecker, J. Kambor and D. Wyler,
\emph{Nucl. Phys. B} {\bf 394} (1993) 101;
J. Kambor, Missimer and D. Wyler,
 \emph{Nucl.\ Phys.\  B} {\bf 346} (1990) 17;
G. Esposito-Far\`ese, 
\emph{Z. Phys. C} {\bf 50} (1991) 255.

\bibitem{EPR87b}
G. Ecker, A. Pich and E. de Rafael, 
\emph{Nucl.\ Phys.  B} {\bf 291} (1987) 692;
 \emph{Phys.\ Lett.\  B} {\bf 237} (1990) 481.

\bibitem{KMW90}
J. Kambor,  Missimer and D. Wyler,
\emph{Nucl. Phys. B} {\bf 346} (1990) 17;
\emph{Phys. Lett. B} {\bf 261} (1991) 496.

\bibitem{PPS00}
E. Pallante and A. Pich,
\emph{Phys.\ Rev.\ Lett.}  {\bf 84} (2000) 2568
  {\tt [hep-ph/9911233]};
\emph{Nucl.\ Phys.\  B} {\bf 592} (2001) 294
  {\tt [hep-ph/0007208]};
 E. Pallante, A. Pich and I. Scimemi,
\emph{Nucl.\ Phys.\  B} {\bf 617} (2001) 441
  {\tt [hep-ph/0105011]}.

\bibitem{CG00}
V. Cirigliano and E. Golowich,
\emph{Phys.\ Rev.\  D} {\bf 65} (2002) 054014
  {\tt [hep-ph/0109265]};
  \emph{Phys.\ Lett.\  B} {\bf 475} (2000) 351
  {\tt [hep-ph/9912513]};
V. Cirigliano, J.F. Donoghue and E. Golowich,
\emph{Eur.\ Phys.\ J.\  C} {\bf 18} (2000) 83
  {\tt [hep-ph/0008290]};
\emph{Phys.\ Rev.\  D} {\bf 61} (2000) 093001
  [Erratum-ibid.\  D {\bf 63} (2001) 059903]
  {\tt [hep-ph/9907341]}.

\bibitem{CEI04}
V. Cirigliano {\it et al.},
 \emph{Eur.\ Phys.\ J.\  C} {\bf 33} (2004) 369
  {\tt [hep-ph/0310351]};
 \emph{Phys.\ Rev.\ Lett.}  {\bf 91} (2003) 162001
  {\tt [hep-ph/0307030]};
G. Ecker {\it et al.},
 \emph{Nucl.\ Phys.\  B} {\bf 591} (2000) 419
  {\tt [hep-ph/0006172]}.

\bibitem{BGP04}
J. Bijnens, E. G\'amiz and J. Prades,
\emph{Nucl. Phys. B (Proc. Suppl.)} {\bf 133} (2004) 245
{\tt [hep-ph/0309216]}.

\bibitem{BB03}
J. Bijnens and F. Borg, 
\emph{Eur.\ Phys.\ J.\  C} {\bf 40} (2005) 383
  {\tt [hep-ph/0501163]};
 ibid. {\bf 39} (2005) 347
  {\tt [hep-ph/0410333]};
{\emph Nucl.\ Phys.\  B} {\bf 697} (2004) 319
  {\tt [hep-ph/0405025]};
J. Bijnens, P. Dhonte and F. Borg,
 \emph{Nucl.\ Phys.\  B} {\bf 648} (2003) 317
  {\tt [hep-ph/0205341]}.

\bibitem{GPS03}
E. G\'amiz, J. Prades and I. Scimemi,
\emph{JHEP} {\bf 10} (2003) 042
{\tt [hep-ph/0309172]};
\emph{Nucl. Phys. B (Proc. Suppl.)}  {\bf 164} (2007) 79
{\tt [hep-ph/0509346]};
{\tt hep-ph/0410150};
 {\tt hep-ph/0405204};
{\tt hep-ph/0305164}.

\bibitem{BP99}
J. Bijnens and J. Prades,
\emph{JHEP} {\bf 01} (1999) 023
  {\tt [hep-ph/9811472]};
  ibid. {\bf 01} (2000) 002
  {\tt [hep-ph/9911392]}.

\bibitem{HPR03}
T. Hambye, S. Peris and E. de Rafael,
\emph{JHEP} {\bf 05} (2003) 027
  {\tt [hep-ph/0305104]};
S. Peris, 
\emph{Nucl.\ Phys. B (Proc.\ Suppl.)} {\bf 133} (2004) 239
  {\tt [hep-ph/0310063]};
{\tt hep-ph/0411308}.

\bibitem{ENJL}
J. Bijnens, C. Bruno and E. de Rafael,
   \emph{Nucl.\ Phys.\  B} {\bf 390} (1993) 501
  {\tt [hep-ph/9206236]};
J. Prades, 
\emph{Z.\ Phys.\  C} {\bf 63} (1994) 491
  [Erratum \emph{Eur. Phys J. C}  {\bf 11} (1999) 571]
  {\tt [hep-ph/9302246]};
J. Bijnens, E. de Rafael and H. Zheng,
\emph{Z.\ Phys.\  C} {\bf 62} (1994) 437
  {\tt [hep-ph/9306323]};
J. Bijnens and J. Prades, 
\emph{Phys.\ Lett.\  B} {\bf 320} (1994) 130
  {\tt [hep-ph/9310355]};
\emph{Z.\ Phys.\  C} {\bf 64} (1994) 475
  {\tt [hep-ph/9403233]};
\emph{Nucl.\ Phys. B\ (Proc.\ Suppl.)}  {\bf 39BC} (1995) 245
  {\tt [hep-ph/9409231]};
J. Bijnens, 
\emph{Phys.\ Rept.}  {\bf 265} (1996) 369
  {\tt [hep-ph/9502335]}.

\bibitem{CDH01}
 J.F. Donoghue and E. Golowich,
  \emph{Phys. Lett. B}  {\bf 478} (2000) 172
  {\tt [hep-ph/9911309]};
V. Cirigliano {\it et al.},
 ibid. {\bf 522} (2001) 245
  {\tt [hep-ph/0109113]};
   ibid. {\bf 555} (2003) 71
  {\tt [hep-ph/0211420]}.

\bibitem{BGP01} 
J.Bijnens, E. G\'amiz and J. Prades,
\emph{JHEP} {\bf 10} (2001) 009
{\tt [hep-ph/0108240]};
\emph{Nucl.\ Phys.\ B (Proc.\ Suppl.)}  {\bf 121} (2003) 195
{\tt [hep-ph/0209089]}.

\bibitem{FGR04}
S. Friot, D. Greynat and  E. de Rafael,
\emph{JHEP} {\bf 10} (2004) 043
{\tt [hep-ph/0408281]};
M. Knecht, S. Peris and E. de Rafael,
\emph{Phys.\ Lett.\  B} {\bf 508} (2001) 117
{\tt [hep-ph/0102017]};
ibid.  {\bf 457} (1999) 227
  {\tt [hep-ph/9812471]}.

\bibitem{NAR01}
S. Narison, \emph{Nucl.\ Phys.\  B} {\bf 593} (2001) 3
  {\tt [hep-ph/0004247]}.

\bibitem{BP00}
J. Bijnens and J. Prades,
  \emph{JHEP} {\bf 06} (2000) 035
  {\tt [hep-ph/0005189]};
  {\tt hep-ph/0009155};
  {\tt hep-ph/0009156};
 \emph{Nucl.\ Phys. B (Proc.\ Suppl.)}  {\bf 96} (2001) 354
  {\tt [hep-ph/0010008]}.

\bibitem{lattice}
 R. Babich {\it et al.},
  \emph{Phys.\ Rev.\  D} {\bf 74} (2006) 073009
  {\tt [hep-lat/0605016]};
 P. Boucaud {\it et al.}
  \emph{Nucl.\ Phys.\  B} {\bf 721} (2005) 175
  {\tt [hep-lat/0412029]};
J.I. Noaki {\it et al.}, [CP-PACS Collaboration],
\emph{Phys. Rev. D} {\bf 68} (2003) 014501
{\tt [hep-lat/0108013]};
T. Blum {\it et al.}, [RBC Collaboration],
\emph{Phys.\ Rev.\  D} {\bf 68} (2003) 114506
  {\tt [hep-lat/0110075]};
D. Be\'cirevi\'c {\it et al.}, [SP$_{QCD}$R Collaboration],
\emph{Nucl.\ Phys. B (Proc.\ Suppl.)}  {\bf 119} (2003) 359
  {\tt [hep-lat/0209136]}.

\bibitem{slopes}
J.R. Batley {\it et al.},  [NA48/2 Collaboration],
 \emph{Phys.\ Lett.\  B} {\bf 649} (2007) 349
  {\tt [hep-ex/0702045]};
 G.A. Akopdzhanov {\it et al.},
  \emph{JETP Lett.}  {\bf 82} (2005) 675
  [Pisma Zh.\ Eksp.\ Teor.\ Fiz.\  {\bf 82} (2005) 771]
  {\tt [hep-ex/0509017]};
I.V. Ajinenko {\it et al.},
  \emph{Phys.\ Lett.\  B} {\bf 567} (2003) 159
  {\tt [hep-ex/0205027]}.

\bibitem{previous}
B.R. Holstein, \emph{Phys. Rev.} {\bf 177} (1969) 2417;
L.-F. Li and L. Wolfenstein, \emph{Phys. Rev. D} {\bf 21} (1980) 178; 
C. Avilez, ibid. {\bf 23} (1981) 1124;
B. Grinstein, S.-J. Rey and M.B. Wise, ibid. {\bf 33} (1986) 1495;
 G. D'Ambrosio {\it et al.},
\emph{Phys. Lett. B} {\bf 273} (1991) 497;
G. Isidori {\it et al.},
\emph{Nucl. Phys. B} {\bf 381} (1992) 522;
 A.A. Bel'kov {\it et al.},
  \emph{Phys.\ Lett.\  B} {\bf 300} (1993) 283;
ibid. {\bf 232} (1989) 118;
 E.P. Shabalin,
  \emph{Phys.\ Atom.\ Nucl.}  {\bf 61} (1998) 1372
  [Yad.\ Fiz.\  {\bf 61} (1998) 1478];
ibid. {\bf 68} (2005) 88
  [Yad.\ Fiz.\  {\bf 68} (2005) 89].

\bibitem{AIM00}
 G. D'Ambrosio, G. Isidori and G. Martinelli,
  \emph{Phys.\ Lett.\  B} {\bf 480} (2000) 164
  {\tt [hep-ph/9911522]}.

\bibitem{NA48asym}
E. Goudzovski, these proceedings;
 J.R. Batley {\it et al.},  [NA48/2 Collaboration],
 {\tt arXiv: 0707.0697 [hep-ex]};
\emph{Phys.\ Lett.\  B} {\bf 638} (2006) 22
  [Erratum-ibid.\  B {\bf 640} (2006) 297]
  {\tt [hep-ex/0606007]};
ibid. {\bf 634} (2006) 474
  {\tt [hep-ex/0602014]}.

\bibitem{ISTRAasym}
 G.A. Akopdzhanov {\it et al.},
\emph{Eur.\ Phys.\ J.\  C} {\bf 40} (2005) 343
  {\tt [hep-ex/0406008]}.


\bibitem{EDI98}
G. Ecker, G. D' Ambrosio, G. Isidori, and  J. Portol\'es, 
 \emph{JHEP} {\bf 08} (1998) 004
{\tt [hep-ph/9808289]}.

\bibitem{GW80}
F.J. Gilman and M.B. Wise,
\emph{Phys. Rev. D} {\bf 21} (1980)  3150.

\bibitem{BLM94} 
A.J. Buras, {\it et al.}, 
  \emph{Nucl.\ Phys.\  B} {\bf 423} (1994) 349
{\tt [hep-ph/9402347]}.

\bibitem{BNL99} 
R. Appel {\it et al.},  [E865 Collaboration],
  \emph{Phys.\ Rev.\ Lett.}  {\bf 83} (1999) 4482
  {\tt [hep-ex/9907045]};
C. Alliegro {\it et al.},
  ibid.   {\bf 68} (1992) 278.

\bibitem{HYP02}
 H. Ma {\it et al.}, [E865 Collaboration],
ibid. {\bf 84} (2000) 2580 {\tt [hep-ex/9910047]};
 H.K. Park {\it et al.}, [HyperCP Collaboration], 
\emph{Phys. Rev. Lett.} {\bf 88} (2002) 111801 
{\tt [hep-ex/0110033]}.


\bibitem{BP93} 
C. Bruno and J. Prades, 
\emph{Z. Phys. C} {\bf 57} (1993) 585
{\tt [hep-ph/9209231]}.

\bibitem{DG95}
J.F. Donoghue and F. Gabbiani,
  \emph{Phys.\ Rev.\  D} {\bf 51} (1995) 2187
  {\tt [hep-ph/9408390]}.

\bibitem{DBI03}
G. D'Ambrosio, G. Buchalla and G. Isidori,
\emph{Nucl.\ Phys.\  B} {\bf 672} (2003) 387
{\tt [hep-ph/0308008]}.

\bibitem{BAT03}
  J.R. Batley {\it et al.}  [NA48/1 Collaboration],
  \emph{Phys.\ Lett.\  B} {\bf 576} (2003) 43
  {\tt [hep-ex/0309075]}.

 \bibitem{BAT04}
J.R. Batley {\it et al.}  [NA48/1 Collaboration],
  \emph{Phys.\ Lett.\  B} {\bf 599} (2004) 197
  {\tt [hep-ex/0409011]}.

\bibitem{ISU04}
G. Isidori, C. Smith and R. Unterdorfer,
\emph{Eur.\ Phys.\ J.\  C} {\bf 36} (2004) 57
{\tt [hep-ph/0404127]}.

\bibitem{FGR05}
S. Friot, D. Greynat and E. de Rafael,
\emph{Phys.\ Lett.\  B} {\bf 595} (2004) 301
{\tt [hep-ph/0404136]};
S. Friot and D. Greynat, 
{\tt hep-ph/0506018}.

\bibitem{MTS06}
F. Mescia, S. Trine and C. Smith,
\emph{JHEP} {\bf 08} (2006) 088
  {\tt [hep-ph/0606081]}.

\bibitem{MS07}
F. Mescia and C. Smith, {\tt arXiv:0705.2025 [hep-ph]}.

\bibitem{BLA07}
  C. Tarantino, these proceedings;
C. Smith, these proceedings;
M. Blanke {\it et al.},
\emph{JHEP} {\bf 06} (2007) 082
  {\tt [arXiv:0704.3329 [hep-ph]]};
   \emph{JHEP} {\bf 01} (2007) 066
{\tt [hep-ph/0610298]};
C. Promberger {\it et al.},
  \emph{Phys.\ Rev.\  D} {\bf 75} (2007) 115007
  {\tt [hep-ph/0702169]};
G. Isidori {\it et al.}
  \emph{JHEP} {\bf 08} (2006) 064
  {\tt [hep-ph/0604074]};
 A.J. Buras {\it et al},
  \emph{Nucl.\ Phys.\  B} {\bf 678} (2004) 455
 {\tt [hep-ph/0306158]}.

\bibitem{BCE99}
J. Bijnens, G. Colangelo and G. Ecker,
  \emph{JHEP} {\bf 02} (1999) 020
  {\tt [hep-ph/9902437]};
\emph{Annals Phys.}  {\bf 280} (2000) 100
  {\tt [hep-ph/9907333]}.

\bibitem{BT02}
J. Bijnens and P. Talavera,
\emph{JHEP} {\bf 03} (2002) 046
{\tt [hep-ph/0203049]}.

\bibitem{PDG06}
 W.-M. Yao {\it et al.} [Particle Data Group],
  \emph{J.\ Phys.\ G} {\bf 33} (2006) 1.

\bibitem{CHARGE}
  E. Abouzaid {\it et al.}  [KTeV Collaboration],
 \emph{Phys.\ Rev.\ Lett.}  {\bf 96} (2006) 101801
  {\tt [hep-ex/0508010]};
A. Lai {\it et al.} [NA48 Collaboration],
  \emph{Eur. Phys. J.  C} {\bf 30} (2003) 33.

\bibitem{PRA07}
J. Prades, in preparation.

\bibitem{KPR00}
M. Knecht, S. Peris and E. de Rafael,
  \emph{Nucl.\ Phys.\ B (Proc.\ Suppl.)}  {\bf 86} (2000) 279
{\tt [hep-ph/9910396]};
E. de Rafael, {\tt hep-ph/0109280};
{\tt hep-ph/0210317}.


\bibitem{KTeV04}
A. Alavi-Harati {\it et al.}  [KTeV Collaboration],
 \emph{Phys.\ Rev.\ Lett.}  {\bf 93} (2004) 021805
  {\tt [hep-ex/0309072]}.

\bibitem{RAG07}
M. Raggi,  
\emph{Nucl. Phys. B (Proc. Suppl.)}  {\bf 167} (2007) 39.

\bibitem{Kpigamma}
J. Bijnens, G. Ecker and A. Pich,
 \emph{Phys.\ Lett.\  B} {\bf 286} (1992) 341
  {\tt [hep-ph/9205210]};
G. Ecker, H. Neufeld and A. Pich,
\emph{Nucl.\ Phys.\  B} {\bf 413} (1994) 321
  {\tt [hep-ph/9307285]};
G. D'Ambrosio and G. Isidori,
  \emph{Z.\ Phys.\  C} {\bf 65} (1995) 649
  {\tt [hep-ph/9408219]};
L. Cappiello and G. D'Ambrosio,
  \emph{Phys.\ Rev.\  D} {\bf 75} (2007) 094014
  {\tt [hep-ph/0702292]}.

\bibitem{BGL03}
J. Bijnens, E. G\'amiz, E. Lipartia, and  J. Prades,
 \emph{JHEP} {\bf 04} (2003) 055
{\tt [hep-ph/0304222]}.

\bibitem{CIR06}
V. Cirigliano {\it et al.},
  \emph{Nucl.\ Phys.\  B} {\bf 753} (2006) 139
  {\tt [hep-ph/0603205]}.

\bibitem{IMT06}
G. Isidori, G. Martinelli and P. Turchetti,
\emph{Phys.\ Lett.\  B} {\bf 633} (2006) 75
  {\tt [hep-lat/0506026]}.

\end{thebibliography}
\end{document}